\documentclass[]{aa}
\usepackage{natbib}
\usepackage{graphicx,xspace}
\usepackage{natbib}
\usepackage{url}
\usepackage{txfonts}
\usepackage{threeparttable}
\usepackage{amsmath, amssymb}
\usepackage{color}
\usepackage{microtype}
\usepackage{threeparttable}
\usepackage{booktabs, makecell}
\usepackage{subfig}
\usepackage{mwe}
\usepackage{subcaption}
\usepackage{lscape}

\def\xmmn{{XMM-Newton}\xspace}

\def\hua{{HU Aqr}\xspace}

\def\msun{M$_\odot$}
\def\na1{Na{\sc I}}
\def\rwd{R$_{\rm WD}$}

\begin{document}

\title{Unveiling the white dwarf in the eclipsing polar HU Aquarii}

\author{A.D. Schwope\inst{1}
\and T.R. Marsh\inst{2}
\and S.G. Parsons \inst{3}
\and J. Vogel\inst{1,4}
\and V.S. Dhillon\inst{3,5}
}

\institute{Leibniz-Institut für Astrophysik Potsdam (AIP), An der Sternwarte 16, 14482 Potsdam, Germany \email{aschwope@aip.de}
\and
Department of Physics, University of Warwick, Gibbet Hill Road, Coventry CV4 7AL, UK
\and
Astrophysics Research Cluster, School of Mathematical and Physical Sciences, University of Sheffield, Sheffield S3 7RH, UK
\and
Department of Visual and Data-Centric Computing, Zuse Institute Berlin (ZIB), Takustr. 7, 14195 Berlin, Germany
\and
Instituto de Astrof\'{i}sica de Canarias, E-38205 La Laguna, Tenerife, Spain
}

\date{Received 18/09/2025; accepted 28/11/2025}

\abstract{We present an analysis of high-speed $u$- and $r$-band photometry of the eclipsing polar \hua  that was obtained with ULTRACAM mounted on the VLT. The observations were performed during a low state, permitting us for the first time to determine the contact points of the white dwarf. Using LCURVE we could determine its size, and hence mass, with a direct method and with unprecedented accuracy.  We determined the mass of the white dwarf as $0.78 \pm 0.02$\,\msun, the mass ratio $Q= M_{\rm WD} / M_\mathrm{sec} = 4.59$,  and the orbital inclination $i=87\fdg4\pm0\fdg9$. An extended warm region with a central temperature of $\sim$33,000\,K was observed in the $u$-band at the location of the previous high-state accretion spot. Weak accretion was ongoing in the low state that led to cyclotron emission that could best be studied with the $r$-band data. It has a diameter of only 3\degr\ to 4\degr\ and is located much closer to the binary meridian than the accretion-heated region studied in the $u$-band. The longitudinal shift of the two accretion regions is of order 30\degr, due to early and late coupling of accreted matter onto the magnetic field lines in low and high accretion states, respectively. The low-state cyclotron-emitting region has a vertical extent of $0.005 - 0.016$\,\rwd, a value that seems to be correlated to the instantaneous accretion rate. }


\maketitle

\begin{figure*}[t]
\begin{center}
\resizebox{\hsize}{!}{\includegraphics[clip=]{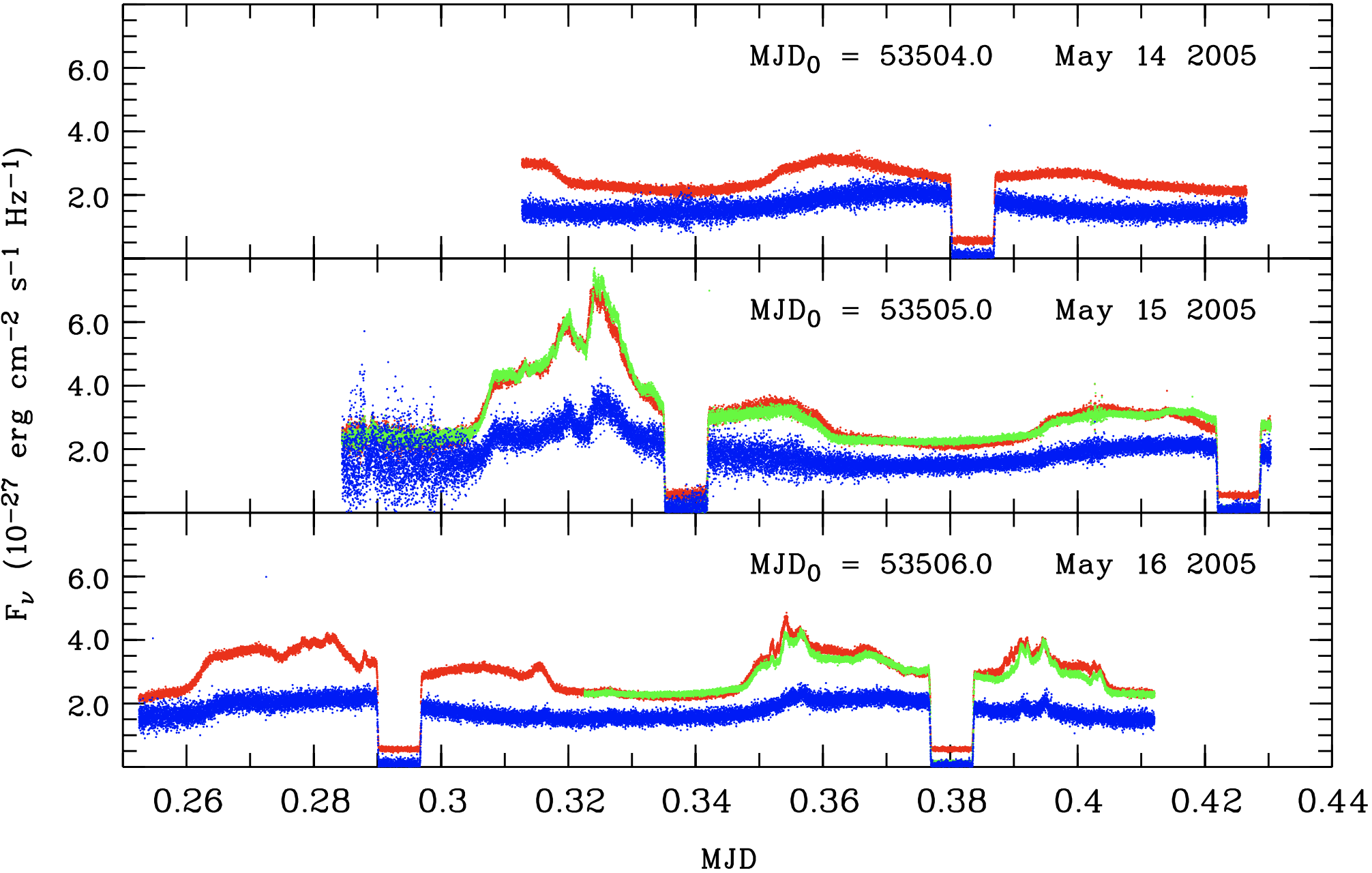}}
\end{center}
 \caption{
  ULTRACAM light curves of HU Aquarii in $u$(blue), $g$ (green), and $r$ (red) in original  time sequence obtained on 14 - 17 May 2005. The data from 17 May are not shown here because just the eclipse was observed.}
\label{f:ulcs}
\end{figure*}

\begin{table*}
\centering
\caption{Observation log of data used in this study.}
\label{t:log}
\begin{tabular}{lclcrrrc}
\hline\hline
Instrument/Telescope & Date & Filter & Resolution & t$_{\rm{start}}$ & t$_{\rm{stop}}$ & t$_{\rm{obs}}$ & runs\\
             & [D/M/Y]           &            & [s]  & [UT]   & [UT]    & [min] \\
\hline
ULTRACAM/VLT & 17/05/2005 & u,g,Na{\sc I} & 0.5 & 9:53 & 10:27 & 34 & 33 \\  
ULTRACAM/VLT & 16/05/2005 & u,g,r \& u,g,He{\sc II} & 0.5  & 6:03   & 9:53    & 230 & 20,21,22\\
ULTRACAM/VLT & 15/05/2005 & u,g,r         & 0.5  & 6:49   &10:19    & 210 & 39,40,41 \\
ULTRACAM/VLT & 14/05/2005 & u,r,He{\sc II}& 0.5  & 7:30   &10:14    & 164 & 13,14\\
\xmmn/OM     & 16/05/2005 & UVM2          & 0.5  & 4:50   &10:22    & 332\\
HST/FOS      & 30/08/1996 &               & 2.5  & 21:27  & 5:51    & 504\\
\hline
\end{tabular}
\tablefoot{Given are the instrument names, the dates, the filters used, the time resolution, and the observation length at the given dates. The last column identifies the ULTRACAM runs (see Sect.~\ref{s:data}).}
\end{table*}

\section{Introduction}
Polars are magnetic cataclysmic variables (CVs) consisting of a late-type main-sequence star and an accreting, synchronously rotating, highly magnetic white dwarf (see \cite{1995cvs..book.....W} for details). The late-type star loses matter via Roche-lobe overflow. After initially following a ballistic trajectory, the matter is caught by the magnetic field of the white dwarf and is funneled onto the surface of the white dwarf where the accretion energy is released in the form of optically thin thermal plasma radiation at X-ray wavelengths and cyclotron radiation, which is detected from IR to UV wavelengths. The photosphere of the white dwarf surrounding the accretion region (or regions) is exposed to this energy and may be reprocessed and reradiated in the extreme ultraviolet and soft X-ray regime.

One of the marked properties of the polars (or AM Herculis stars according to the prototype) is the occurrence of high and low states of accretion. Mass loss changes from the donor star might happen in other CV types as well, but are more difficult to recognize if there is an accretion disk acting as a mass storage.\footnote{A notable exception are the VY Scl stars, disk-accreting Novalikes, which may show pronounced low states} In the absence of a disk, changes in the donor star mass loss rate immediately lead to changes in the accretion rate onto the white dwarf. Low states occur randomly; the accretion duty cycle is of order 50\% in the prototype AM Herculis and only 5\% in the high-field system AR UMa \citep{hessman+00, mason+24}, but is not well characterized for a representative sample of objects. Due to the largely reduced accretion rate during such states, all sources of accretion-induced radiation are efficiently dimmed and the stellar components of the binary can  be observed more easily. 

\hua is one of the most intensively studied polars because it is the brightest eclipsing object \cite[maximum orbital brightness $V_{\rm max} \simeq 14.5$][]{schwope+01}, and thus offers enormous diagnostic potential. Since its discovery in 1992 (\citealt{1993MNRAS.263...61H}, \citealt{schwope+93}), it was observed in all accessible wavelength ranges, mostly during states of high or intermediate accretion. It has a one-pole accretion geometry with a field strength of 34 MG \citep{schwope+03}. The accretion region shows self-eclipses by the white dwarf, which makes it possible to discern between the bright and faint phases.

The observer and the accretion spot lie in the same hemisphere with respect to the orbital plane. Hence, the accretion stream and accretion curtain guided by magnetic field lines give rise to pronounced absorption features in optical and X-ray light curves and occur 0.08 - 0.15  orbital phase units prior to the eclipse \citep{schwope+01}.

In this paper we present an analysis of high-speed photometric observations of \hua obtained with ULTRACAM \citep{2007MNRAS.378..825D} when it was attached to one of the Very Large Telescope (VLT) UTs. Simultaneous observations with \xmmn were performed, and the results of the X-ray observations were reported by \cite{schwarz+09}. Some of the results of the ULTRACAM observations were used in other studies \citep{vogel+08, schwope+11}, but the data themselves were never properly published. 
We are taking the opportunity of the colloquium held in Warwick in September 2024 celebrating Tom Marsh's legacy after his untimely passing to correct this omission \citep{pelisoli+25}.

\section{Observations}
\label{section_obs}
HU Aquarii was observed with the high-speed photometer ULTRACAM \citep{2007MNRAS.378..825D} mounted on the 8.2m Very Large Telescope at the Paranal Observatory, Chile, during four nights between 14 May and 17 May 2005 (see Table \ref{t:log} for a log of all observations used in this study). ULTRACAM is a triple-beam camera that can record images in three filters simultaneously. The observations were performed through Sloan $u$,\,$g$,\, and $r$ filters and a filter centered on C{\sc III}/N{\sc III} + He{\sc II} (central wavelength 4662\,\AA, FWHM 108\,\AA). The $u$ and $r$ filters were used throughout the whole campaign, while the the $g$ and He{\sc II} filters were used alternatively. The time resolution achieved was 0.5 seconds with only 24 msec of dead time between exposures. The data were reduced with the ULTRACAM pipeline. The comparison star for differential photometry is indicated on the finding chart in \cite{schwope+93}. 

During the last night of the campaign (16--17 May) simultaneous X-ray and UV-observations were performed with \xmmn. The X-ray data were described in detail in \cite{schwarz+09} and are not further addressed here.

\hua was encountered in a low state of accretion in May 2005. Figure \ref{f:ulcs} shows all ULTRACAM data in the continuum filters obtained during the mentioned four nights. The mean brightness during the faint phase was $u = 18\fm5$, $g = 18\fm0$, and $r = 18\fm1$. On the night of 14 May, the bright phase appeared at its faintest level with the $u$-band light curve modulated only by the foreshortening of the heated region surrounding the former accretion spot. Its maximum occurred at MJD = 53504.3735 (orbital phase $\sim$0.87). During the two following nights some flare-like brightness enhancements were observed affecting all filters. The $r$-band light curve on 14 May shows a double-humped modulation similar to the high state, where it was explained by beaming of the cyclotron radiation \citep[see][their Fig.~2]{schwope+03} and indicates some weak residual accretion. The low-state $r$-band humps from Fig.~\ref{f:ulcs}, upper panel, occur at MJD 53504.3632 and 53504.4005, which correspond to orbital phases $-0.25$ and $+0.18$, respectively. The same explanation, modulation of the $r$-band light curve by cyclotron radiation, is adopted here. The pre-eclipse dip due to absorption of light in the magnetically guided accretion stream, a remarkable feature in the high-state light curves, was not observed in the low state of May 2005. In the high state, it was centered at orbital phase $\sim$0.87, which corresponds to MJD = 53504.3736, 53505.3287, 53505.4155, 53506.2836, and 53506.3703, respectively, for the ULTRACAM observations.

\begin{figure*}
\begin{center}
\includegraphics[width=\columnwidth,clip=]{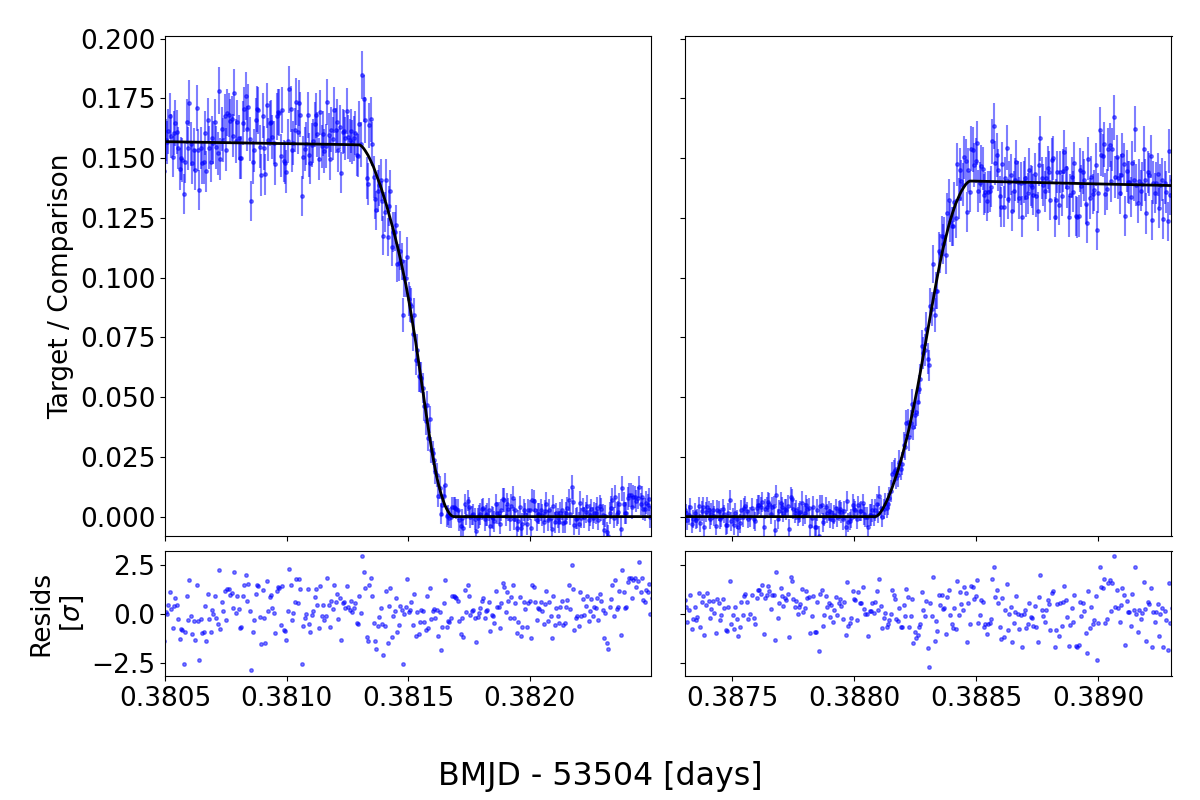}
\includegraphics[width=\columnwidth,clip=]{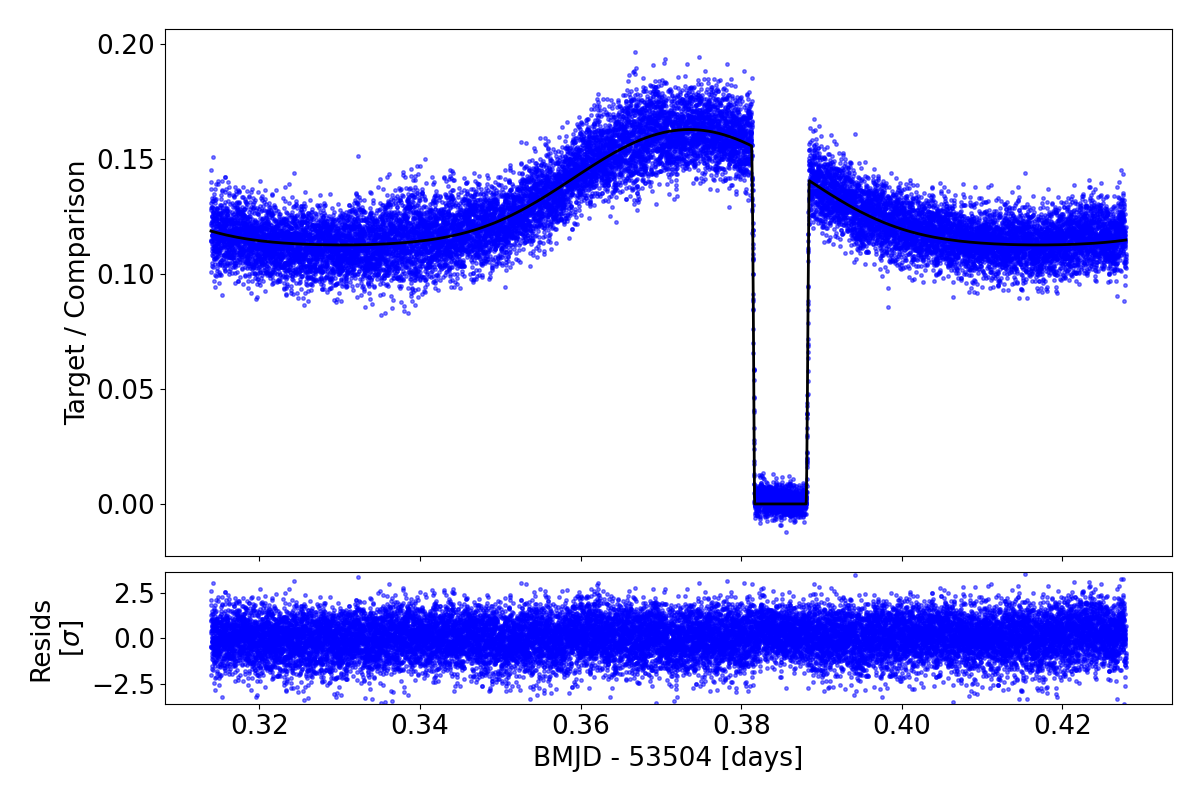}
\end{center}
\caption{ULTRACAM $u$-band light curve from 14 May 2005 with model fit overplotted. The left panels show the fit to the ingress and egress of the white dwarf eclipse, while the right panel shows the fit to the full light curve. The residuals to the fit are shown at the bottom (in standard deviations).}
\label{f:ubandfit}
\end{figure*}

\begin{figure}
\begin{center}
\includegraphics[width=\columnwidth,clip=]{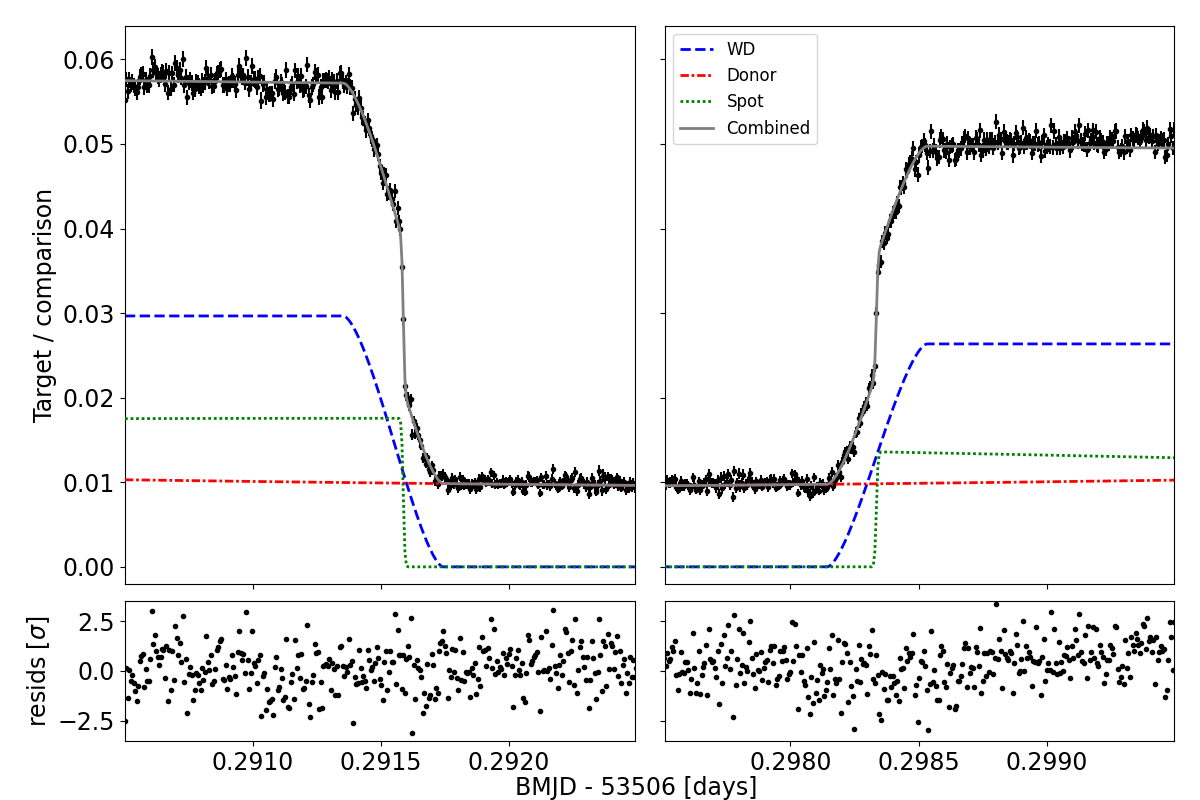}
\end{center}
\caption{ULTRACAM $r$-band eclipse light curve from 16 May 2005 with best-fit model shown in gray. The individual components of the model are also shown: white dwarf (blue dashed line), donor star (red dot-dashed line), cyclotron spot (green dotted line).  The residuals are shown below (in standard deviations).}
\label{f:n3ecl}
\end{figure}

\begin{figure}
\begin{center}
\includegraphics[width=\columnwidth,clip=]{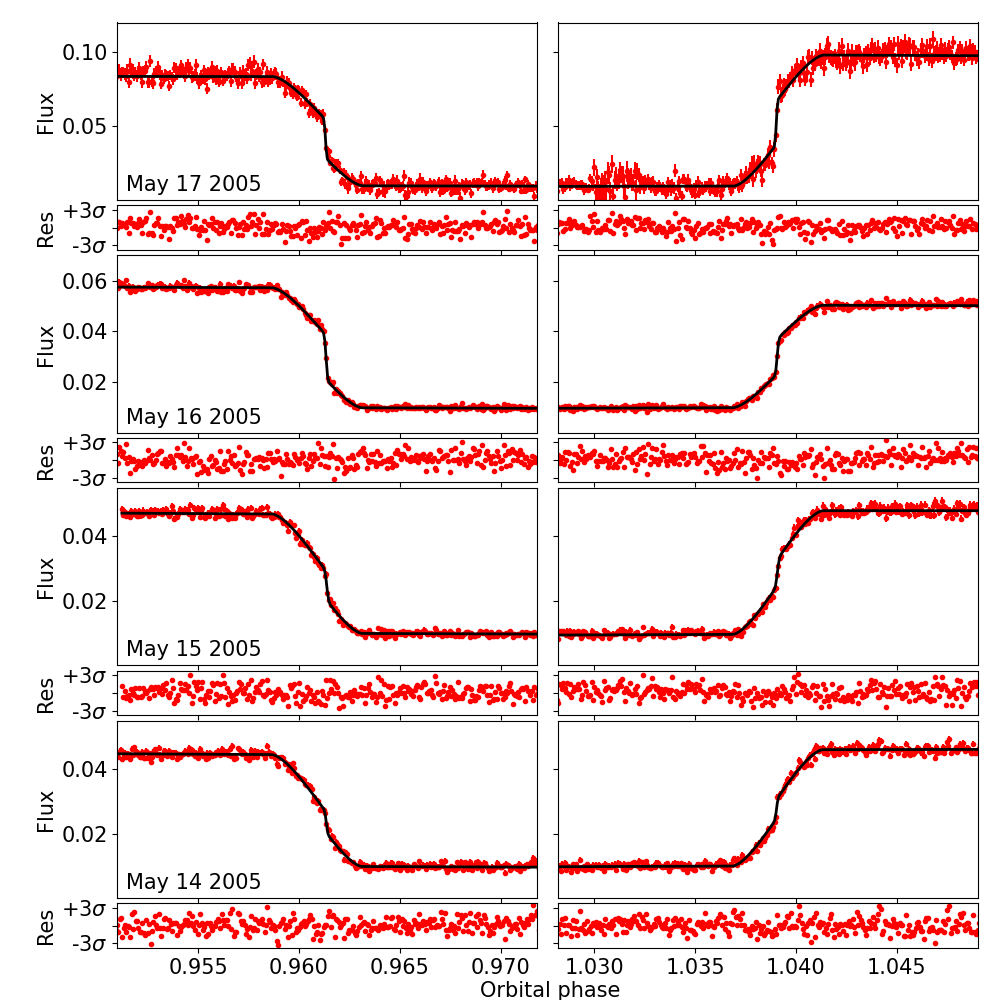}
\end{center}
\caption{ULTRACAM $r$-band eclipse light curves with model fits overplotted. The fit includes the eclipse of both the white dwarf photosphere and cyclotron spot. The residuals are shown below each fit (in standard deviations).}
\label{f:recl}
\end{figure}

The eclipse light curves obtained in the $u$- and $r$-bands show similarities, but also some remarkable differences (see Figs.~\ref{f:ubandfit} and \ref{f:n3ecl}) that give us the opportunity to measure the size of the white dwarf and some parameters of the heated region where accretion happened in the preceding high accretion state, and to study the location and dimension of the actual accretion region.

\section{Analysis: LCURVE modeling of ULTRACAM high-speed photometry}

We modeled the ULTRACAM light curves using LCURVE \citep{2010MNRAS.402.1824C}, a code written specifically to model the light curves of compact binary systems containing at least one white dwarf. LCURVE can simulate many of the features seen in the light curves of HU Aqr, including eclipses, Roche distortion, and some accretion phenomena. It is only able to model nonmagnetic accretion, although it is possible to add hot spots to the surface of the white dwarf to mimic some effects of magnetic accretion. Fortunately, many of the ULTRACAM observations caught the system in an extremely low state which allowed  us to model features that would usually be too distorted by accretion variability to fit reliably.

The eclipse of the white dwarf in the $u$-band is particularly clean, and all four of the contact points are clearly visible. We therefore modeled these data with LCURVE to constrain the stellar and binary parameters of HU Aqr. In particular, we focused on the data from the first night (14 May 2005) where the accretion rate appears to be at its lowest. LCURVE requires many input parameters to create a model light curve and for this fit the crucial ones are the mass ratio $q=M_\mathrm{WD}/M_\mathrm{sec}$, the orbital inclination $i$, the radii of both stars scaled by the orbital separation $R_\mathrm{WD}/a$ and $R_\mathrm{sec}/a$, the temperatures of the two stars $T_\mathrm{WD}$ and $T_\mathrm{sec}$, the limb darkening coefficients for both stars, the orbital period $P_\mathrm{orb}$, and the time at the center of the eclipse $T_0$. 

\begin{table}[t]
\centering
\caption{Best-fit parameters for the $u$-band light curve.\label{t:ufit}}
\tabcolsep=1.5mm
\centering
\begin{tabular}{lccc}
\hline\hline
Parameter & Unit & Value & Prior \\
\hline
$M_\mathrm{WD}$ & $M_\odot$ & $0.78 \pm 0.02$ & U(0, 1.4) \\
$M_\mathrm{sec}$ & $M_\odot$ & $0.17 \pm 0.01$ & U(0, 1.0) \\
Inclination $i$ & deg & $87.4 \pm 0.9$ & U(0, 90) \\
$T_0$-53504 & TDB & 0.3848869(18) & U(0.3, 0.4) \\
Spot longitude $\chi$ & deg & $46.9 \pm 0.3$ & U(-180, 180) \\
Spot latitude $\delta$ & deg & $67.2^{+3.4}_{-4.1}$ & U(-90, 90) \\
Spot FWHM & deg & $44.0^{+6.5}_{-5.9}$ & U(0, 180) \\
Spot $T_\mathrm{cen}$ & K & $33150^{+8900}_{-5900}$ & U(0, $10^5$) \\
\hline
\end{tabular}
\tablefoot{Best-fit parameters from fitting the ULTRACAM $u$-band light curve of HU Aqr from 14 May 2005. We note that $T_0$ corresponds to the center of the white dwarf eclipse, not the mid-egress time as is more typically used for HU Aqr. It is given in BJMD. The last column gives the range of the prior; flat priors were assumed for all parameters.}
\end{table}

When fitting the $u$-band data from 14 May 2005, we kept $P_\mathrm{orb}$ fixed at 0.0868203980 days from \citet{bours+14} \citep[for a complete history and latest update of the ephemeris see ][]{schwope_thinius18}. 
Since a single $u$-band eclipse is insufficient to constrain the temperature of the white dwarf $T_\mathrm{WD}$, it was fixed at 13,500 K \citet{vogel+08}. These authors used a code similar to LCURVE, but less advanced, to model the ULTRACAM and the simultaneous XMM-OM data. The derived temperature was similar to that derived by \cite{gaensicke99} from the light curve models applied to the intermediate-state HST/GHRS data; they derived 14,000\,K. We note that LCURVE assumes blackbody temperatures for all components, and hence these are essentially flux scaling factors rather than actual temperatures. After redetermining the white-dwarf radius and mass from the $u$-band eclipse profile we checked the results for consistency with respect to the white-dwarf temperature (see below). We also fixed $T_\mathrm{sec}$ to 3000 K, although this makes no difference to the fit whatsoever because the donor star flux is not detected in the $u$-band.

In addition to these fixed values we forced the donor star radius to be the same as the Roche lobe radius. We also forced the white dwarf to follow the theoretical mass-radius relationship of \citet{2020ApJ...901...93B} for a 13,500 K white dwarf. For the limb darkening coefficients, we interpolated the four-parameter nonlinear limb darkening values from \citet{2020A&A...634A..93C} for the white dwarf and fixed the limb darkening of the donor star to a linear value of 0.5. Since the donor star is not detected in the $u$-band this value has no effect on the fit.

\begin{figure}[t]
\begin{center}
\includegraphics[width=\columnwidth,clip=]{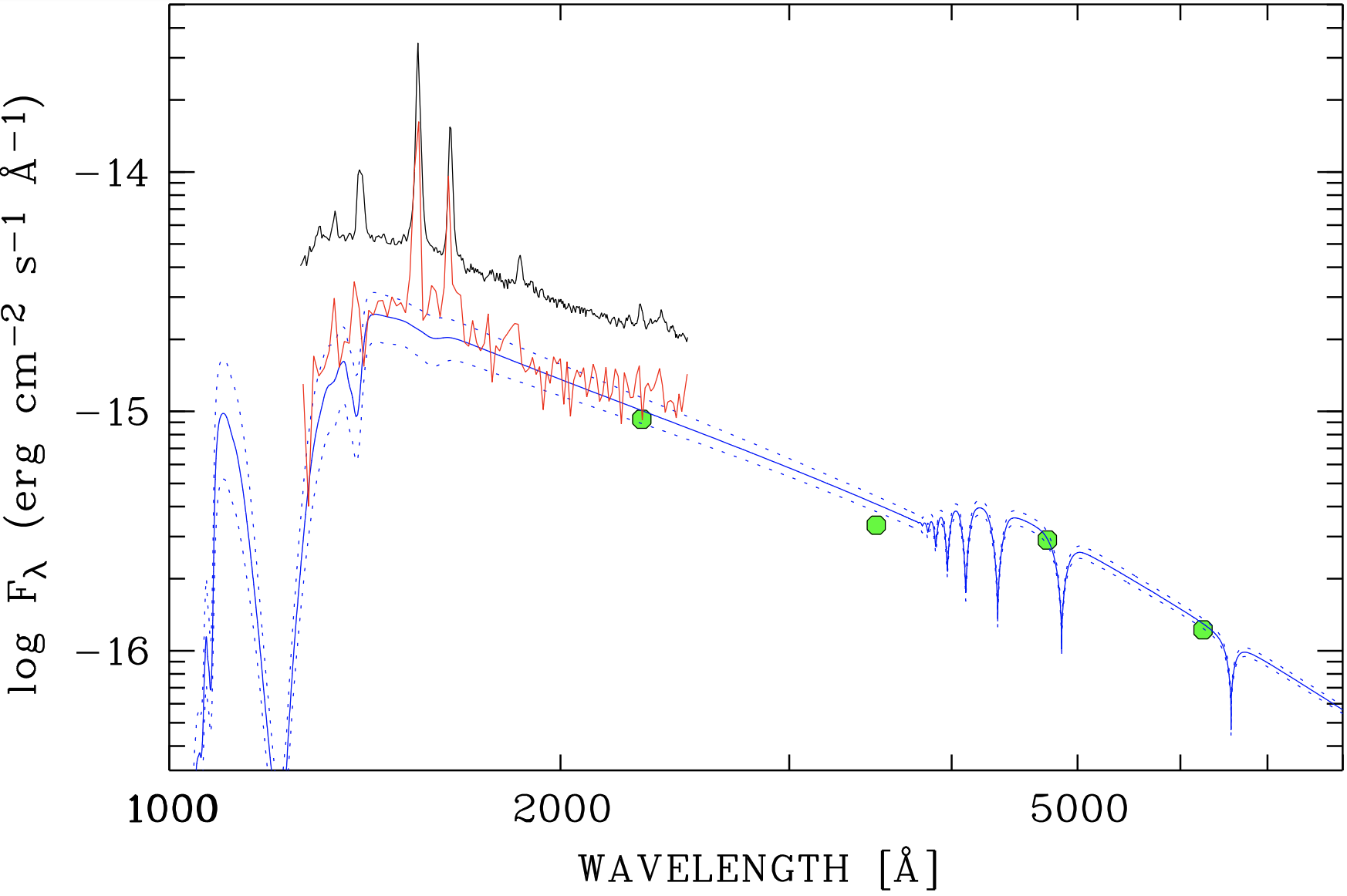}
\end{center}
\caption{UV to optical spectral energy distribution of \hua. Shown are the faint-phase data only, the accretion steam-corrected HST/FOS spectrum obtained 1996, and the OM data obtained simultaneously with the ULTRACAM $u,g,r$ photometry. The model spectra for 13,500 K, 14,000 K, and 14,500 K scaled to the distance and radius of the white dwarf in \hua are shown as blue lines.}
\label{f:sed}
\end{figure}

Finally, in order to properly model the $u$-band light curve we also needed to account for the warm spot on the surface of the white dwarf from the former accretion spot. The spot was modeled with a Gaussian distribution of temperatures with the central temperature and full width at half maximum (FWHM) as free parameters. The longitude $\chi$ (relative to the meridian defined by the line of centers of the two stars) and the latitude $\delta$ (relative to the orbital plane, or colatitude $\beta = 90\degr - \delta$ relative to the rotation axis) of the spot are also free parameters, and we assumed that the white dwarf rotates at the orbital period.

The fit was performed using a custom Python code and the Markov chain Monte Carlo (MCMC) method, as implemented in the emcee Python package \citep{2013PASP..125..306F}. The input parameters to the fit were the masses of the two stars $M_\mathrm{WD}$ and $M_\mathrm{sec}$, the inclination $i$, the time of superior conjunction of the white dwarf $T_0$,  and the spot longitude, latitude, FWHM, and central temperature. At each step in the fit the input parameters are used to generate an LCURVE model. The white dwarf mass is combined with the mass-radius relationship to calculate the white dwarf radius, and both masses are used with Kepler's third law to compute the orbital separation. The ratio of the masses also sets the Roche lobe of the donor and hence its radius. This method allowed us to break the well-known degeneracy between the (scaled) radii of the two stars and the orbital inclination when fitting just the white dwarf eclipse data \citep{2017MNRAS.470.4473P}. Uniform priors were placed on all the input parameters to keep them within the physical limits (see Table~\ref{t:ufit}). For the fit we used 100 walkers, with a burn-in period of 1500 steps and 10000 production steps. 

The fit to the $u$-band data from 14 May 2005 is shown in Fig.~\ref{f:ubandfit} and the corner plot that was used to determine the error ranges is shown in the Appendix (see Fig.~\ref{f:ucorn}). The best-fit parameters with their uncertainties are listed in Table~\ref{t:ufit}. These values give a time interval of 33.6 seconds between the first and second (and third and fourth) white-dwarf eclipse contact points, a total eclipse duration (first to fourth contact points) of 620.3 seconds, and an eclipse duration for the center of the white dwarf of 586.7 seconds. The best-fit binary parameters of Table \ref{t:ufit} are in excellent agreement with previous studies of HU Aqr \citep[e.g.,][]{schwope+11}, but with improved error ranges. The uncertainties in $i$, $\chi$, and $\delta$ became unprecedentedly small. The longitude $\chi$ is fixed by  the eclipse details and the orbital phase of maximum brightness, while $i$ and $\delta$ are fixed by the detailed shape of the eclipse. The only significant assumption made here is that the white dwarf follows the mass-radius relationship of \citet{2020ApJ...901...93B}. We therefore chose to use these stellar and binary parameters when modeling the more complex $r$-band light curves.

The best-fit parameters are in agreement, but nevertheless slightly different compared to earlier studies. The earlier studies had to assume a distance to \hua, meanwhile a precise distance is available from Gaia astrometry. We therefore tested if the assumed temperature of 13,500 K together with the radius of the WD and the distance to \hua fit well with its ultraviolet to optical spectral energy distribution. For this we used the average faint-phase HST/FOS spectrum obtained in an intermediate state of accretion  in 1996 \citep{schwope+04b}, corrected for the emission from the accretion stream and curtain, the faint-phase optical monitor (OM) brightness from the simultaneous ULTRACAM/XMM-Newton observations \citet{schwarz+09}, and the ULTRACAM $u,g,r$ faint-phase data. These are shown in Fig.~\ref{f:sed} together with the WD model atmosphere spectra for 13,500 K, 14,000 K, and 14,500 K, respectively \citep{koester10}. This comparison suggests $T_{\rm eff, WD} = 13,500 - 14,000$\,K, and thus confirms our initial assumption. For completeness we ran a fit to the $u$-band data with $T_{\rm eff, WD}$ fixed at 14,000\,K, which gave the same results within the errors as the fit for 13,500\,K, whose parameters are reported in Table \ref{t:ufit}.

Having established the stellar and binary parameters (and by extension the contact points of the white dwarf eclipse), we then constrained the location of the cyclotron component that is seen to eclipse in the $r$-band. The offset between the contact points of the white dwarf and cyclotron component eclipses can be used to determine the location of the latter.

We modeled the $r$-band eclipses with LCURVE taking a similar approach to the $u$-band fit, but with some notable differences. First, we performed fits to the $u$-band eclipse data from all nights using the model shown in Fig.~\ref{f:ubandfit} and only allowing the central eclipse time ($T_0$) to vary. We then fitted the $r$-band eclipse data for each night. For these fits we kept the two masses and inclination (hence scaled radii) fixed at the values given in Table~\ref{t:ufit}. The central eclipse time ($T_0$) was allowed to vary, but we placed a Gaussian prior on this value based on the value and uncertainty from fitting the $u$-band data from the same eclipse. We kept the spot component on the white dwarf, but allowed its longitude, latitude, FWHM, and central temperature to vary during the fit. 

The $r$-band data also show considerably more out-of-eclipse variability than the $u$-band, likely related to accretion not being accurately modeled by LCURVE. Therefore, we limited our fit to just the ingress and egress phases to avoid needing to account for this extra variability. However, even with limiting the fit to just these phases, there is still variability that is not accounted for in the LCURVE model. In particular, the pre- and post-eclipse flux levels are generally different due to cyclotron beaming and short-term changes of the accretion rate;  some eclipses are brighter before the eclipse and some are brighter after the eclipse. To try to account for this we allowed independent scaling factors for the ingress and egress. LCURVE uses temperatures as scale factors, so we allowed the temperature of the white dwarf to be different during the ingress and egress, essentially allowing the additional emission on one side of the eclipse to be included in the white dwarf contribution. While this is clearly not physical, the main aim of these fits was to accurately model the contact points of the eclipse; we found that this approach allowed the depth of the ingress and egress to be different, which is vital in order to accurately measure the contact points. The fits were performed using the MCMC method with 100 walkers, a burn-in period of 2000 steps, and 10000 production steps. 

\begin{figure*}
\begin{center}
\includegraphics[width=\textwidth,clip=]{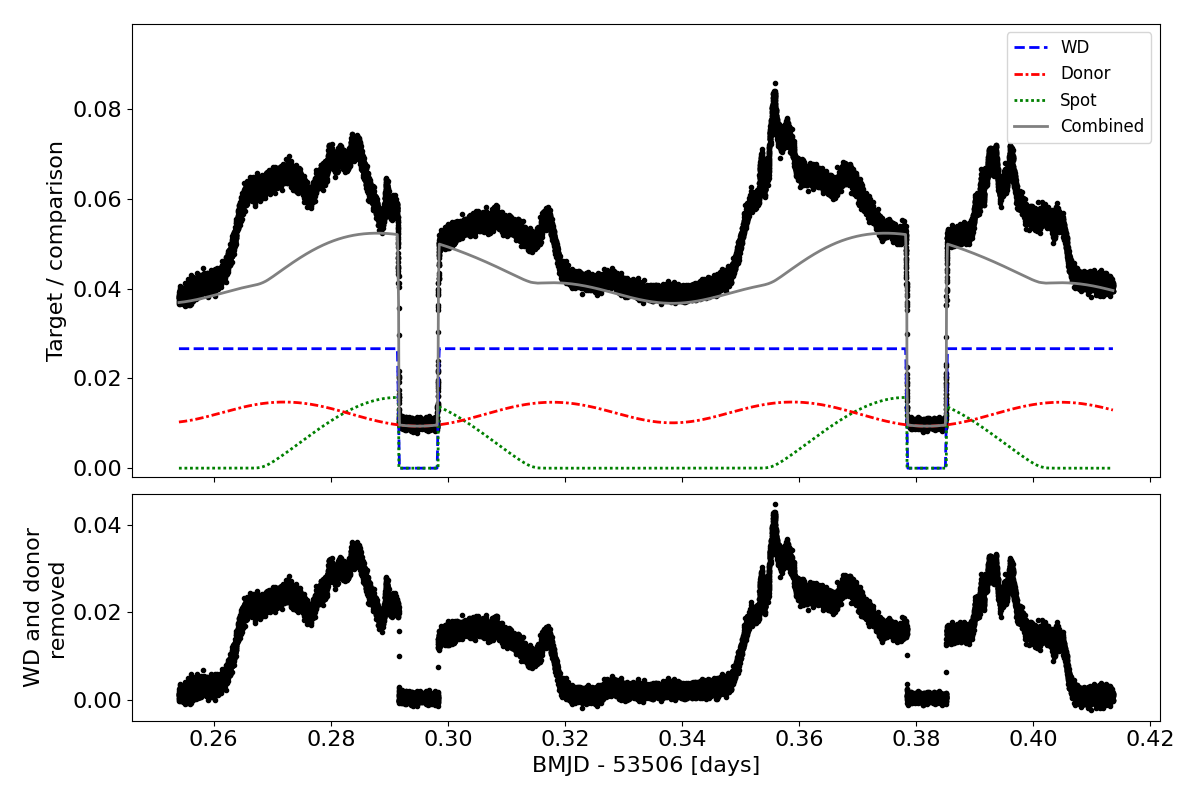}
\end{center}
\caption{Top panel: ULTRACAM $r$-band light curve from 16 May 2005 with the best-fit model to the eclipse phases shown in gray (and extended to the full orbit). The individual components of the model are also shown: white dwarf  (blue dashed line), donor star (red dot-dashed line), cyclotron spot (green dotted line). Bottom panel: Light curve with the white dwarf and donor star contributions removed, thus mainly showing cyclotron emission.}
\label{f:n3lc}
\end{figure*}

\begin{table*}
\centering
\caption{Best-fit parameters for the $r$-band light curve \label{t:rfit}}
\begin{tabular}{lccccc}
\hline\hline
Parameter & Unit & 14 May 2005 & 15 May 2005 & 16 May 2005 & 17 May 2005 \\
\hline
Spot longitude & deg & $15.3^{+0.6}_{-0.5}$ & $16.9\pm0.4$ & $14.6\pm0.2$ & $12.4\pm0.6$ \\
Spot latitude & deg & $69.0^{+0.2}_{-0.3}$ & $68.3\pm0.2$ & $67.3\pm0.1$ & $68.2\pm0.3$ \\
Spot FWHM & deg & $3.2^{+1.0}_{-0.7}$ & $4.2\pm0.4$ & $3.4\pm0.1$ & $2.9\pm0.3$ \\
Spot $T_\mathrm{cen}$ & $10^6$ K & $2.9^{+0.7}_{-0.6}$ & $2.2^{+0.5}_{-0.3}$& $5.8\pm0.4$ & $10.1^{+2.1}_{-1.6}$ \\
\hline
\end{tabular}
\tablefoot{Best-fit parameters for the cyclotron emission region based on fitting the ULTRACAM $r$-band light curves of HU Aqr. We note that the temperature of the spot is purely a flux scaling factor and assumes blackbody emission, which is not an accurate representation of cyclotron emission.}

\end{table*}

\begin{figure}
\begin{center}
\includegraphics[width=\columnwidth,clip=]{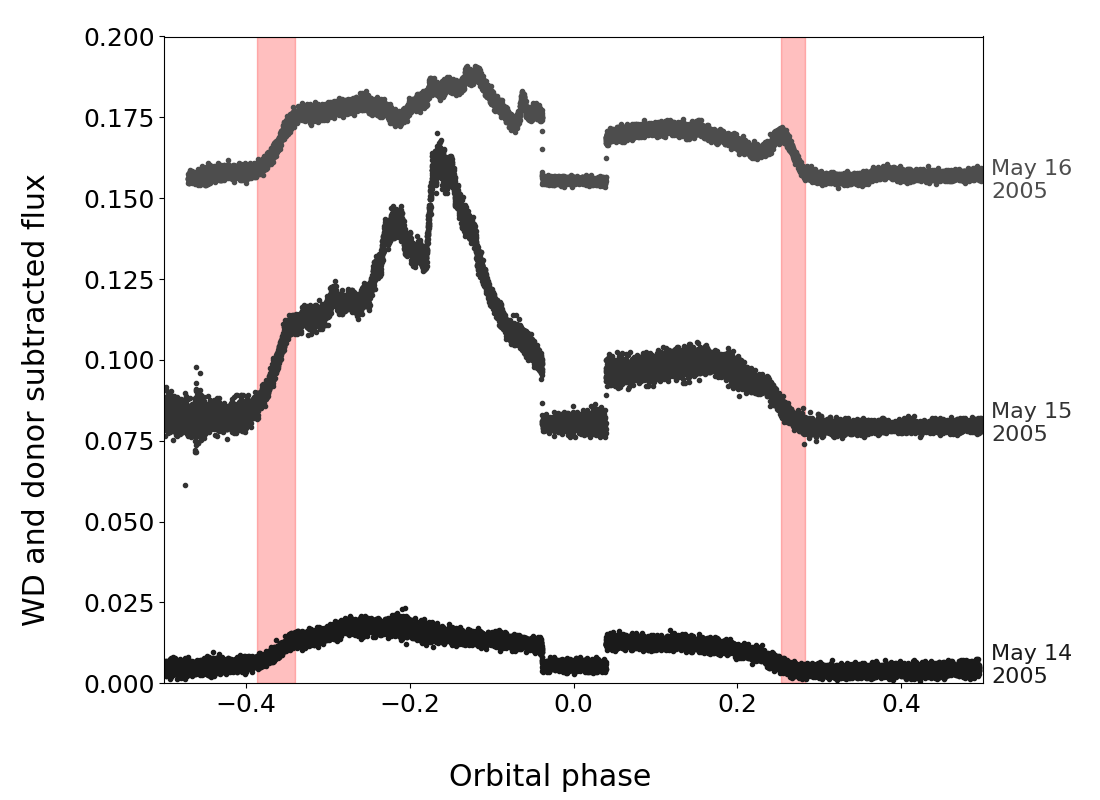}
\end{center}
\caption{Three ULTRACAM $r$-band light curves of HU Aqr with the white dwarf and donor star contributions subtracted. Each curve is offset vertically by 0.075 for clarity. The shaded pink regions indicate the approximate phases where the additional variability appears and disappears on 16 May 2005.}
\label{f:rres}
\end{figure}

An example best fit is shown in Fig.~\ref{f:n3ecl} for an eclipse on 16 May 2005; the individual components of the fit are also shown. It can be seen that the white dwarf component needs to be brighter before the eclipse, which is likely not the case in reality, but is the result of an additional accretion flux pre-eclipse that the model compensates for by making the white dwarf component brighter. Fits to the $r$-band eclipses from all four nights are shown in Fig.~\ref{f:recl} and the measured spot parameters are listed in Table~\ref{t:rfit}. The corner plot used to derive uncertainties for the night 3 fit is shown in the Appendix (Fig.~\ref{f:rcorn}). 
Taken at face value, the numbers given in Table~\ref{t:rfit} indicate a small, but significant (4$\sigma$), shift of the cyclotron emission region from night to night. LCURVE models the spots simply as a hotter (hence brighter) region on the surface of the white dwarf with the flux from the spot changing purely due to the viewing angle and eclipse. However, cyclotron photons are emitted perpendicular to magnetic field lines with some beaming patterns, so a simple viewing angle approximation may not necessarily be an accurate representation. Moreover, LCURVE does not take into account any vertical extent to the cyclotron emission region (it assumes that all emission comes from the surface of the white dwarf), which could affect these values, a scenario that is tested further below. Time-dependent accretion and the variable vertical extent may lead to some scatter in the derived spot location, and  the shift in the spot longitude and latitude might be apparent only.

The temperature of the spot (essentially a measure of how bright it is) clearly changes from night to night. In the first night it is very weak and faint, in the last night the spot eclipse is very clear and deep in the light curve. Despite all these caveats we note that the cyclotron radiation is emerging from a very small region with a lateral extent of only 3\degr\ to 4\degr. 

In Fig.~\ref{f:n3lc} we show the full ULTRACAM $r$-band light curve from 16 May 2005. We use the best-fit model to the eclipse (using the egress scale factors to ensure  not to overcorrect the post-eclipse flux) and extended it over the full data range. We also show the model broken  into its individual components. The white dwarf gives a constant flux except when it is eclipsed, while the donor star shows a double-peaked light curve as a result of the Roche distortion. Subtraction of the contributions of the donor, the white dwarf, and the spot gives a light curve that may be regarded as  pure cyclotron radiation. This is shown in the bottom panel of Fig.~\ref{f:n3lc}. The eclipse of the cyclotron spot is clearly visible, as is a considerable amount of additional variability through nonstationary, time-dependent accretion when the spot is visible.

Figure~\ref{f:rres} shows the white dwarf and donor-subtracted $r$-band light curves for several nights. All of the bright phases are thought to be powered mainly by cyclotron radiation from the small accretion spots (for the extent, see line 3 in Table~\ref{t:rfit}). They show the same overall characteristics. There is an almost linear increase into the bright phase and on its descent (indicated by the shaded bands). The bright phase itself is modulated by cyclotron beaming (best visible on 14 May through its double-humped light curve pattern) and/or by instantaneous changes in the mass accretion rate. This behavior dominates on 15 and 16 May, when the overall mass accretion rate was enhanced. The small measured extent of the accretion spot that emits cyclotron radiation allows us to approximate it as a point on the surface of the white dwarf. The length of the self-eclipse of the accretion spot, $\Delta\phi_{\rm self}$; the orbital inclination, $i$; and the colatitude $\beta = 90\degr - \delta$ are related via $\cos{(\pi\Delta\phi_{\rm self}}) = \tan^{-1}(i) \, \tan^{-1}(\beta)$. This relation predicts a length of the self-eclipse of 0.464 phase units, which is much longer than the observed value, or a length of the bright phase of 0.536 phase units, which is much shorter than the observed value. The observed length of the self-eclipse is approximately 0.34 phase units. Taking this at face value, and assuming a spot on the surface of the white dwarf, a short self-eclipse can be realized if the latitude of the spot is of order 85\degr. This high latitude does not conform to the precisely measured ingress and egress phases of the spot ($\phi_{\rm ing} = -0.038642\pm 0.000027, \phi_{\rm egr} = -0.039047\pm 0.000017$, spot eclipse length $582.74\pm 0.23$, s), which fix the latitude at approximately 68\degr\ with very high precision and reliability. The remedy is to assume a vertical extent of the cyclotron emission region. 

We tested this scenario with a code similar to LCURVE. It has the same geometric functionality, but is less elaborate with respect to the physics of emission and the light curve of the donor, for example. It was used successfully in an earlier study of soft X-ray eclipses in HU Aqr \citep{schwope+01}. The emission is thought to originate from a cylindrical column with a radius that corresponds to the measured lateral extent of the accretion spot (Table~\ref{t:rfit}). The column has a certain height and is divided into 100 equidistant layers. Each layer has the same number of tiles. For an assumed height, the code tests under Roche geometry which tiles are visible at any given phase. We used the same period; the same masses, and hence mass ratio; and the same white dwarf radius as used for the LCURVE models. We  only asked for the number of tiles that are visible at any given phase and did not consider any foreshortening or beaming. The output is normalized to unity. The height is modified until an agreement is reached between the observed and modeled lengths of the bright phases. 

\begin{figure}[t]
\begin{center}
\resizebox{\hsize}{!}{\includegraphics[clip=]{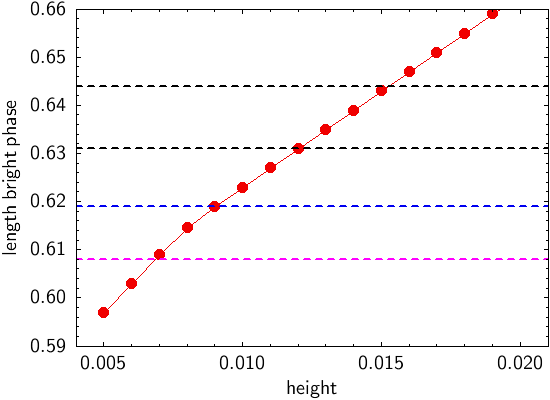}}
\end{center}
\caption{Predicted vs. observed length of the bright phase (FWHM) for a column on the white dwarf at a latitude of 68\degr and an azimuth of 18\degr as a function of its height. The height is given in units of white-dwarf radii. The chosen inclination is $87\fdg4$. The dashed horizontal lines illustrate the observed lengths of the four fully covered bright phases on 14 May (magenta), 15 May (blue), and 16 May (black). }
\label{f:hei}
\end{figure}

The results of this exercise are illustrated in Fig.~\ref{f:hei}. The modeled and observed lengths of the bright phase are shown as a function of the column height (in units of white-dwarf radii). Four complete bright phases were covered, and their individual lengths were determined as the phase interval between half light during their initial ingress and egress phases. The modeled length was simply the FWHM of the bright phase. The comparison shows that the column height evolved from 0.006 \rwd\ at the lowest state on 14 May to 0.015 \rwd\ on 16 May, when the overall brightness of \hua had increased considerably.   

\section{Discussion and conclusion}
\label{s:disc} 

We presented an analysis of high-speed photometry with ULTRACAM when it was mounted at the VLT. The target was the bright eclipsing polar \hua. The details of the eclipses in the $u$- and $r$-bands were studied and modeled with LCURVE. The object was encountered in an unusually low state of accretion. Due to this, the white dwarf could be uniquely located in the binary system for the first time. From modeling the data obtained on 14 May in the lowest accretion state, the white-dwarf mass was inferred from the measured size (length of eclipse ingress and egress) and found to be 0.78\,\msun, close to the value of  $\sim$0.8\,\msun\ determined as the median mass in CV sample studies of their X-ray or UV-spectra \citep{shaw+20, pala+22}. 

The $u$-band light curve displays a maximum at orbital phase 0.870. The smoothly variable brightness variation outside the eclipse and the details of the eclipse were both successfully modeled assuming a heated region at a longitude of 47\degr and a latitude of about 67\degr (for the exact values and their uncertainties, see Table~\ref{t:ufit}). The binary inclination was redetermined and found to be $87\fdg4$. Assuming a Gaussian-shaped symmetric heated region, its maximum temperature was 33,000\,K, and its size (FWHM) 44\degr. The large extent gives rise to the smooth brightness variation. The maximum spot temperature is highly dependent on the assumed temperature profile and should not be taken too literally. The measured longitude (azimuth) $\chi$ agrees very well with the longitude of the high-state X-ray emitting accretion spot, while its latitude ($\delta=67.2\degr$, colatitude $\beta=22.8\degr$) is higher by a few degrees compared to the colatitude of 25--31\degr  found in the earlier study. However, at the time of the analysis of the X-ray data, the exact location of the white dwarf in the system was not yet known; it is still difficult or even impossible to fix exactly at that occasion because it is outshined by accretion-induced radiation. Its location affects the derived (co)latitude of the accretion spot. We therefore consider the heated region detected in the ULTRACAM $u$-band as the surroundings of the former high-state soft X-ray emitting region, which had a diameter of about 3\degr\ \citep{schwope+01}. 

The binary parameters found here, which we consider the most precise and reliable set of parameters yet achieved for HU Aqr, are in excellent agreement with a previous determination \citep{schwope+11}, but with considerably smaller uncertainties. The previous study involved Ca{\sc II} emission and Na absorption lines from the irradiated and non-irradiated hemispheres of the donor star, respectively, to constrain the mass ratio. Our new results presented here confirm the validity of the previous approach with the additional benefit of the spectroscopic method that it can also be applied to non-eclipsing objects. 

The $r$-band data obtained simultaneously on 14 May and even more so those obtained during the following nights show a different emission component:  beamed cyclotron radiation from some weak residual accretion. The cyclotron-emitting region is much smaller than the heated region around the former high-state accretion spot. Its location and size were also determined with LCURVE and found to be at the same (co)latitude as the warm region, but shifted by about 30\degr\ in longitude closer to the binary meridian. It has a diameter of only 3\degr\ to 4\degr, i.e., it is of similar size to the high-state soft X-ray emitting spot, but at a different location. Cyclotron radiation was also found in a previous high state of accretion. It was emitted along an accretion arc with a total extent of about 20\degr\ \citep{schwope+03}. The eclipse ingress lasted 4 seconds, the egress 9 seconds. The position of the far end of the arc, as seen from the donor, was well in agreement with the high-state accretion-heated spot studied here based on the ULTRACAM $u$-band data. The near end of the high-state arc was found at a longitude of about 25\degr to 30\degr, farther away from the binary meridian, as determined here from the ULTRACAM $r$-band data. The low-state scatter in the longitude of the spot was between 12\degr\ and 17\degr. The matter that feeds the low-state cyclotron spot is threaded onto magnetic field lines much closer to $L_{\rm 1}$ than in the high state, and it is tempting to assume that matter couples to the field directly at $L_{\rm 1}$.

Although the $r$-band eclipse could be well modeled with LCURVE, the implied length of the bright phase was found to be shorter than observed. The assumption of a vertical extent of the cyclotron-emitting plasma solves the discrepancy. The cyclotron-emitting region has a height that varies between 0.005 and 0.016 white dwarf radii, and is correlated with the overall brightness of the system. The vertical and the lateral extent could both be well determined. The derived values imply that the column (as a synonym for the emission region) is pillbox-shaped rather than pencil-shaped with a height-to-length ratio of about 1/5. Interestingly, both the high-state cyclotron and the soft X-ray emitting regions have a similar height of about 0.015\,\rwd, which means that the correlation between the column height and the accretion rate is valid only for a very restricted range in the low-state accretion rate.

The eclipse length of the cyclotron emitting spot found here is 582.7 s, that of the high-state soft X-ray spot was 587.5 \citep{schwope+01}. Both structures have comparable sizes, the eclipse ingress and egress length of the optical and the egress length of the X-ray eclipse are comparable (the ingress length of the X-ray eclipse could not be measured precisely due to the unknown effect of soft X-ray absorption in the high-state accretion curtain) and the heights are comparable as well. Hence, any difference in the eclipse length seems to be due to a shift in the colatitude, the low-state cyclotron-emitting spot being located at a higher latitude than the high-state X-ray spot. However, there is a caveat. The length of the ROSAT-observed X-ray eclipse is longer than the white-dwarf eclipse determined here, $\Delta T_{\rm X} =587.5$\,s versus $\Delta T_{\rm WD} = 586.7$\,s. This seems counterintuitive because the accretion spot is located in the upper hemisphere; one thus might expect $\Delta T_{\rm X} $ to be shorter than $\Delta T_{\rm WD}$, but the spot lies closer to the obscuring donor than the center of the white dwarf, thus extending the eclipse length. Another effect is that the measurement of the soft X-ray eclipse is biased toward a more extended length because of soft X-ray absorption at the eclipse ingress. One needs a better defined X-ray eclipse to finally determine the latitude of the X-ray emitting spot.

Our study has shown how observations of a well-selected target with a world-leading instrument mounted on a prime facility and analyzed with suitably adapted software gives unprecedented insight into the fine details of accretion onto a strongly magnetic white dwarf. The binary parameters of \hua could ultimately be settled.

\section{Data availability \label{s:data}}
The ULTRACAM data (see Tab.~\ref{t:log}) are available in electronic form at the CDS via anonymous ftp to cdsarc.u-strasbg.fr (130.79.128.5) or via http://cdsweb.u-strasbg.fr/cgi-bin/qcat?J/A+A/.

\begin{acknowledgements}
We acknowledge constructive criticism by an anonymous referee. ULTRACAM and VSD are supported by STFC grant ST/Z000033/1. JV was supported by Deutsches Zentrum f\"ur Luft- und Raumfahrt (DLR) GmbH under contract No.~50 OR 0404.  
\end{acknowledgements}

\bibliographystyle{aa}
\bibliography{000hultra}

\begin{appendix}

\section{Posterior probability distributions for the ULTRACAM $u$- and $r$-band fits}
~~~~~~~~~~
\begin{figure}[h]
\begin{center}
\resizebox{\textwidth}{!}{\includegraphics[clip=]{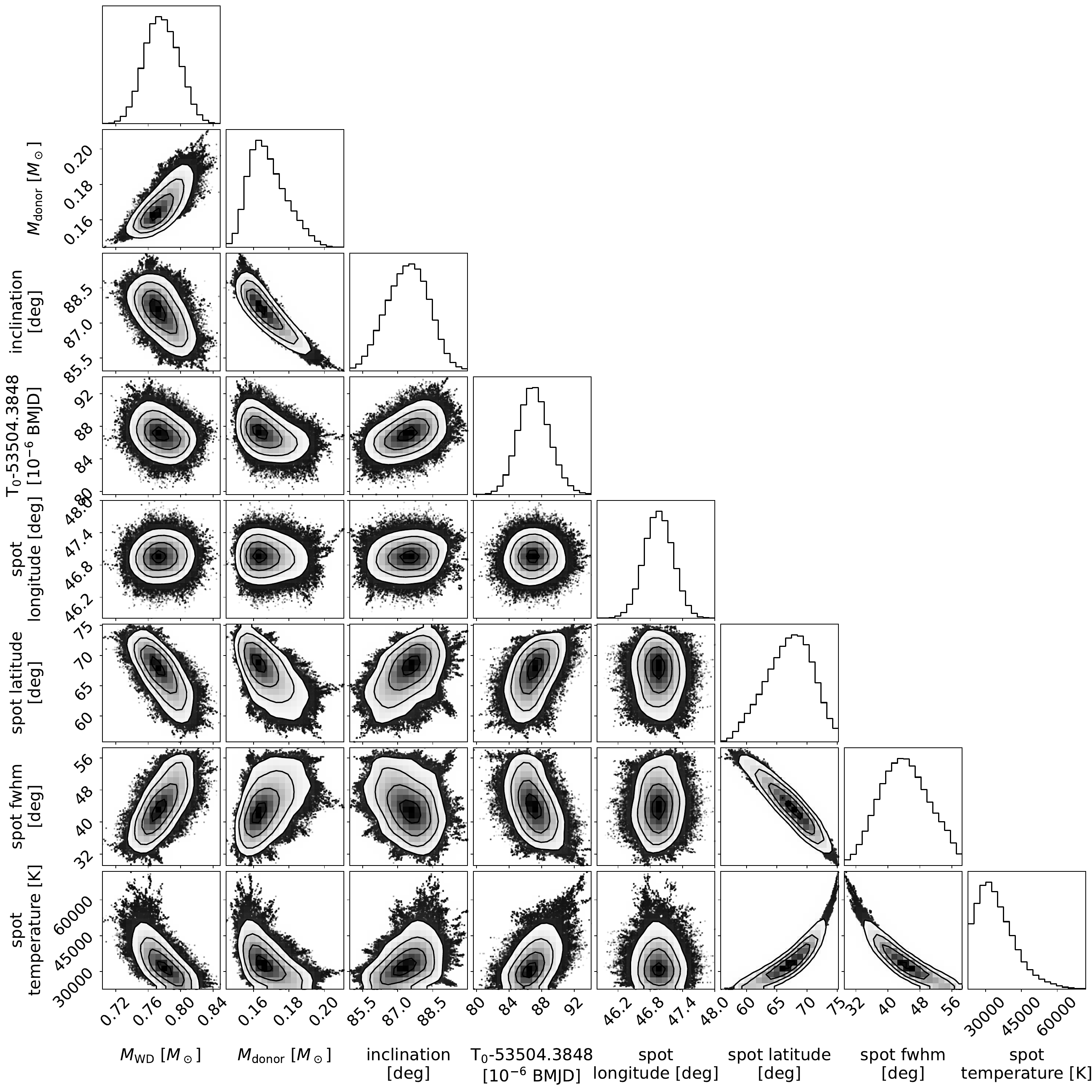}}
\end{center}
\caption{Posterior probability distributions of the seven parameters for the $u$-band fit (see main text for details).}
\label{f:ucorn}
\end{figure}

\begin{figure*}
\begin{center}
\includegraphics[width=\textwidth,clip=]
{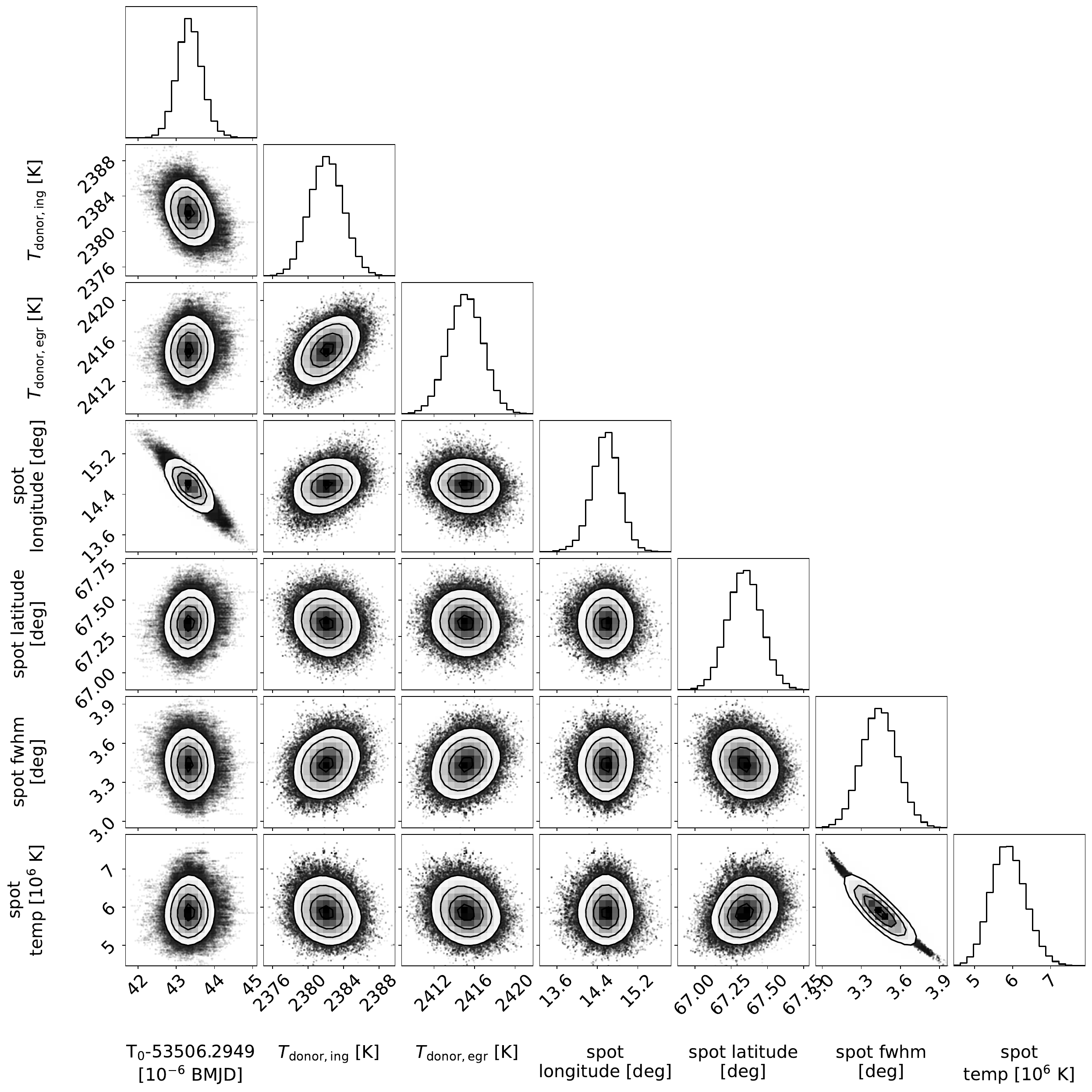}
\end{center}
\caption{Posterior probability distributions of the six parameters for the $r$-band fit of the third night (see main text for details). The size of the white dwarf and its time of superior conjunction were kept fixed during the fit at the values found from the $u$-band fit.}
\label{f:rcorn}
\end{figure*}
\end{appendix}
\end{document}